\documentclass[12pt]{article}
\usepackage{epsfig}
\usepackage{amsfonts}
\usepackage{amsopn}
\usepackage{amsmath}

\title{
\vskip -50 pt
\begin{flushright}
\normalsize\rm AEI-2012-050 \\ 
\end{flushright}
\vskip 20 pt
Matrix regularization of embedded 4-manifolds}

\author{
Maciej Trzetrzelewski \thanks{e-mail: maciej.trzetrzelewski@gmail.com} \\ \\
Max-Planck-Institut f\"ur Gravitationsphysik \\
Albert-Einstein-Institut, \\
M\"uhlenberg 1, D-14476 Potsdam, \\
Germany}

\begin{document}
\date{}
\maketitle

\abstract{We consider products of two $2$-manifolds such as $S^2 \times S^2$, embedded in Euclidean space and show that the corresponding $4$-volume preserving diffeomorphism algebra can be approximated by a tensor product $SU(N)\otimes SU(N)$ i.e. functions on a manifold are approximated by the Kronecker product of two $SU(N)$ matrices. 

A regularization of the $4$-sphere is  also performed by  constructing $N^2\times N^2$  matrix representations of the $4$-algebra (and as a byproduct of the $3$-algebra which makes the regularization of  $S^3$ also possible).}

\section{Motivation}

Matrix regularization of surfaces is a remarkable statement which says that there exists a map from functions on a surface (closed, orientable) to $SU(N)$ matrices such that structure constants of the area preserving diffeomorphisms (APD) on that surface can be approximated by the structure constants of $SU(N)$. The statement was put froward in early 80s by Goldstone and Hoppe \cite{Hoppephd} in the case of a sphere. Later on it was generalized to the case of a torus \cite{torus1,torus2,torus3}, higher genus Riemann surfaces \cite{hgenus1,hgenus2} and arbitrary K\"ahler manifolds \cite{kahler1,kahler2}. A very important property of matrix regularization is that among closed, connected and orientable surfaces the underlying matrix group is $SU(N)$ regardless of the topology. On the other hand, the topological information is encoded in the way one proceeds with the $N \to \infty$ limit therefore the global information about the surface is in the limit. At finite but large $N$ one can extract topological information about the surface by analyzing the spectral properties of the matrices as observed in \cite{Shimida}.

Originally, matrix regularization was invented in order to consistently quantize a membrane \cite{Hoppephd}. However it is clear that the framework itself, a priori has nothing to do with quantization: it is a mathematical framework in which one can consider functions on surfaces as a limiting  case of certain  $N \times N$ matrices. As such it can be applied both to classical physics by considering the regularized classical equations of motion \cite{class1,class2,class3} and quantum physics by regularizing the quantum theory of fields on the surface - as a result obtaining a tractable quantum mechanical system \cite{Hoppephd,quant}. In the last few years it has been discovered that many aspects of the differential geometry of (embedded) Riemannian $2$-manifolds can be formulated by that procedure \cite{hha1,hha2,hha3}.

Matrix description of a quantum membrane is clearly interesting from the particle-physics point of view. If elementary particles are extended objects then it is natural to consider membranes and their excitations as good candidates - the idea put forward by Dirac in 1962 \cite{Dirac1,Dirac2,Dirac3} where he introduced what is now called the Nambu-Goto type action for membranes and considered the possibility that leptons may be understood as excitations of the membrane ground state (the electron). 

On the other hand, it is quite natural to search for generalizations of matrix-like regularization to higher-dimensional manifolds. This problem, although very well known \cite{HoppeT}, is still unsolved. Moreover from the point of view of quantum gravity one is more interested in a possible regularized description of space-time itself which favors $4$-manifolds among others. If, for simplicity, one assumes that the $4$-manifold under consideration is compact and that it has Euclidean signature then one arrives at as simple question as: what is the regularized description of  $S^2\times S^2$? In this paper we shall prove that the group structure underlying this manifold is the tensor product $SU(N)\otimes SU(N)$ i.e. a group of $N^2 \times N^2$ matrices obtained by a Kronecker product of two $SU(N)$ matrices. The result can be generalized to arbitrary product $\mathcal{N}\times \mathcal{M}$ of two $2$-manifolds $\mathcal{N}$ and $\mathcal{M}$ whose regularizations are given by $SU(N)$ matrices.

A way to regularize the $4$-sphere is  also presented by constructing $N^2\times N^2$ matrix representations of the coordinates of $S^4$ embedded in $\mathbb{R}^5$. 
We than show that the Nambu $4$-bracket evaluated on the coordinates can be replaced by a $4$-commutator of those matrices. This construction can then be used to prove analogous statements for $S^3$.

\section{Matrix regularization of surfaces}

Let us start with reviewing matrix regularization for surfaces. There are several excellent reviews on that subject e.g. \cite{Hopperev, Nicolai} - our objective here is to fix the notation and the conventions. 

For the case of a single sphere it is convenient to expand functions on $S^2$ in terms of spherical harmonics $Y^l_m(\varphi_1,\varphi_2)$
\[
F(\varphi_1,\varphi_2) = \sum_{l=0}^{\infty} \sum_{m=-l}^l a^l_m Y^l_m(\varphi_1,\varphi_2)
\]
where $a^l_m$ are real. The corresponding structure constants $c_{ABC}$ of the APD of $S^2$ will be given by
\[
\{Y_A,Y_B\} = c_{ABC}Y_C
\]
where we used a double-index notation $A=(l,m)$ and where $\{\cdot,\cdot\}$ is the Poisson bracket
\begin{equation} \label{def}
\{f,g\}:= \frac{1}{\rho} \left(\partial_{\varphi_1}f \partial_{\varphi_2} g - \partial_{\varphi_2}f \partial_{\varphi_1} g \right), \ \ \ \rho=\sin \varphi_1.
\end{equation}
The factor $ \rho$ is conventional at this stage however it turns out to be a unique one (up to a constant) for which the regularization of the of a membrane Hamiltonian can 
performed. By analyzing membrane equations of motion one finds that, up to a constant, $\rho=\sqrt{\det g_{rs}}$ where $g_{rs}$ is a metric of the embedded surface.

Other reason why that factor is preferred is due to the identity for the coordinates of the unit sphere
\[
\{x_i,x_j\}= \epsilon_{ijk}x_k,
\]
\[
x_1=\cos \varphi_1, \ \ \ x_2 = \sin \varphi_1 \cos \varphi_2, \ \ \ x_3 = \sin \varphi_1 \sin \varphi_2
\] 
i.e. under a Poisson bracket defined as in (\ref{def}) the coordinates of a unit sphere satisfy relations similar to those for the spin matrices. This observation turns out to be  crucial for matrix regularization of $S^2$. Note also that the Poisson bracket satisfies all the axioms of a Lie bracket (i.e. the antisymmetry and the Jacobi identity) hence it provides  the Lie algebraic structure for functions on $S^2$.

Matrix regularization maps spherical harmonics $Y_A$ to matrices  $T_A $ in such a way that when replacing the Poisson bracket $\{\cdot,\cdot\}$ by a commutator $-i[\cdot,\cdot]$ the structure constants $f_{ABC}^{(N)}$ given by $-i[T_A,T_B]=f_{ABC}^{(N)}T_C$  converge in the large $N$ limit to $c_{ABC}$. Let us give more details for that construction for $S^2$.

First step is to use the fact that spherical harmonics can be expressed in terms of homogeneous polynomials in coordinates $x_1$, $x_2$, $x_3$
\[
Y^l_m(\theta,\varphi) =\sum_{i+j+k=l} a^m_{ijk} x_1^i x_2^j x_3^k = \sum_{a_i=1,2,3} c^m_{a_1 \ldots a_l}x_{a_1}\ldots x_{a_l} 
\]
where in the last equality we introduced the coefficients $c^m_{a_1 \ldots a_l}$ which are totally symmetric. The second step is to replace the coordinates $x_i$ by $N\times N$ spin matrices $J_i=S_i/\sqrt{s(s+1)}$, $[S_i,S_j] = i\epsilon_{ijk}S_k$  where $s$ is the spin (in particular $2s=N-1$ and  $s(s+1)=(N^2-1)/4$ is an eigenvalue of the Casimir operator $S^2$)
 obtaining a map
\begin{equation} \label{tas}
Y^l_m \to T^l_m :=\gamma_{lN}  \sum_{a_i=1,2,3} c^m_{a_1 \ldots a_l}J_{a_1}\ldots J_{a_l}
\end{equation}
with the constant $\gamma_{lN}$ to be determined (see below). The choice of spin matrices $J_i$ is motivated by an algebraic identity $\sum_i J_i^2 = \mathbf{1}$ which mimics the corresponding identity for the coordinates $\sum_i x_i^2=1$ (however one can use other conventions e.g. $J_i=S_i$ as in original work by Hoppe \cite{Hoppephd}). At this stage there is no reason yet to think that this choice will give rise to structure constants which in the large $N$ limit converge to the structure constants of APD's of $S^2$. Therefore at this stage it would be premature to refer to the above construction as to the construction of a matrix-regularized $S^2$. 

The key observation is to note that the Poisson bracket on $x_i$'s  has the same structure as when considering  spin matrices with the commutator
\[
\{x_i,x_j\}= \epsilon_{ijk}x_k,
\]
\begin{equation} \label{eq1}
-i[J_i,J_j] = \frac{2}{\sqrt{N^2-1}}\epsilon_{ijk} J_k
\end{equation}
(which was mentioned above) and to observe that both the Poisson bracket and the commutator satisfy the Leibniz rule
\[
\{F,GH\} = \{F,G\}H + F\{G,H\},
\]
\begin{equation} \label{eq2}
-i[\hat{F},\hat{G}\hat{H}] = -i [\hat{F},\hat{G}]\hat{H} -i \hat{F}[\hat{G},\hat{H}]
\end{equation}
for functions $F$, $G$, $H$ and their matrix counterparts $\hat{F}$, $\hat{G}$, $\hat{H}$.
Because of (\ref{eq1}) and (\ref{eq2}), when calculating the structure constants of $T_A$'s by evaluating $[T_A,T_B]$ one will perform the same type of manipulations as if one was calculating $\{Y_A,Y_B\}$ modulo factors of $\frac{1}{\sqrt{N^2-1}} \sim 1/N$ for large $N$. It is the main reason why the structure constants $f_{ABC}^{(N)}$ converge to $c_{ABC}$.
What remains to calculate is the  constant $\gamma_{lN}$ in (\ref{tas}).  It must be fixed so the all the factors of $1/N$ will be canceled when calculating an arbitrary commutation relation $[T_A,T_B]$.  The correct choice turns out to be $\gamma_{lN}\propto N$ for large $N$ with the proportionality factor depending on the conventions (see \cite{Hoppemt} for a detailed discussion). 

\section{General considerations}

To generalize the construction outlined in the previous section, to the case of higher-dimensional manifolds, is a challenging task. Let us point out some of the most important difficulties related to such construction.

\subsection{The $N^d$ scaling}
First, let us note that the number of $N^2-1$ real parameters of the $SU(N)$ matrix is closely related to the topological dimension of the surface i.e. for $d$-dimensional manifold we expect the $SU(N)$ matrices to be replaced by objects parametrized by $\sim N^d$ real numbers, for large $N$. The simplest way to see this is to consider a $d$-dimensional torus $T^d$. Functions on $T^d$ can be expanded in terms of  a convenient basis 
\begin{equation} \label{basist}
e^{i\varphi_1 n_1} \ldots e^{i\varphi_d n_d}, \ \ \ \ \varphi_k \in [0,2\pi)
\end{equation}
where $n_k$'s are integers. Because non of the $\varphi_k$'s are distinguished we expect that the regularization based on the truncation of the mode numbers $n_k$ should be such that $n_k \le N$ for some $N$ and all $k$'s. Therefore there would be $N^d$ parameters corresponding to a single expansion in terms of (\ref{basist}). 

A similar conclusion is obtained when analyzing the expansion on $S^d$ in terms of corresponding $S^d$-spherical harmonics $Y^l_{m_1\ldots m_{d-1}}(\varphi_1,\ldots,\varphi_{d})$. These functions can be classified by homogeneous polynomials of order $l$ of $d+1$ variables  \cite{hypers1,hypers2}. The number of such polynomials for fixed $l$ is given by
\begin{equation}  \label{dld}
D_{l d}=\binom{l+d}{d} - \binom{l+d-2}{d} 
\end{equation}
(e.g. $2l+1$ for $d=2$, $(l+1)^2$ for $d=3$) i.e. it scales with $\frac{2}{(d-1)!}l^{d-1}$ for large $l$. Therefore the sum of the first $N$ modes, for large $N$, gives
\begin{equation} \label{scaling}
\sum_{l=0}^{N-1}D_{l d} \approx \frac{2}{(d-1)!} \sum_{l=0}^{N-1}l^{d-1} \approx \frac{2}{d!}N^{d}.
\end{equation}

Another way to argue the scaling $N^d$ is to view the regularized $d$-manifold as a lattice of interacting points (i.e. vertices and links). Here one immediately arrives at the conclusion that there should be about $N^d$ such points. However this argument is less rigorous then the previous one since already for surfaces it is not clear what is a geometrical meaning of the $N^2-1$ parameters. Moreover as observed by Nicolai and Helling \cite{Nicolai} the matrix regularization is significantly different from lattice approaches as it does not introduce any dimensional parameters (unlike in lattice approaches where one introduces a lattice spacing). 

Finally we note that the scaling $N^d$ (for large $N$) satisfies also consistency checks e.g. for $S^2 \times S^2$ one should obtain $\sim N^4$ real parameters, as for $S^4$ or $T^4$, which is indeed the case - there are $N^2$ modes corresponding to each $S^2$ in $S^2 \times S^2$.

\subsection{$d$-arrays}
The above considerations are simple but suggest something less trivial i.e. that the possible algebraic objects corresponding to functions on $d$-manifolds should be certain generalizations of square matrices. A natural proposal is to consider $d$-dimensional matrices or $d$-arrays i.e. matrices that carry more then two indices which range from $1$ to $N$ and a related concept of $d$-algebras (for a review see \cite{darray}) . There exist however elementary problems with such objects e.g. for $(2n+1)$-arrays  it is difficult to define a product of two arrays. If a product of two arrays involves only index contractions then it is not possible to multiply two $(2n+1)$-arrays in such a way that the result is again the $(2n+1)$-array. For example a product of two $3$-arrays can be a $4$-array a $2$-array or a $0$-array
\[
(\hat{F}\bullet \hat{G})_{ABCD}:=\hat{F}_{ABE}\hat{G}_{ECD}, \ \ \ \  (\hat{F}\bullet \hat{G})_{AB}:=\hat{F}_{ACD}\hat{G}_{DCB}, 
\]
\[
(\hat{F}\bullet \hat{G}):=\hat{F}_{ABC}\hat{G}_{CBA}
\]
but never a $3$-array (here $\bullet$ indicates a product of $d$-arrays, $\hat{F}$ and $\hat{G}$ are $d$-arrays corresponding to functions $F$ and $G$ respectively). To overcome this difficulty one considers a multiplication rule in which some indices are fixed and equal but not summed over \cite{cubic1,cubic2}. Similar problem does not appear for $2n$-arrays i.e. it is possible to define a product of two $2n$-arrays so that the product involves only index contractions and the result is again a $2n$-array. For $4$-arrays an example is
\[
(\hat{F}\bullet \hat{G})_{ABCD}:=\hat{F}_{ABEF}\hat{G}_{FECD}
\]
and clearly there are many other examples of this sort obtained by permuting the indices and considering linear combinations. 

The apparent dichotomy between odd and even dimensions is the reason why regularization of $d$-manifolds should be carried out in a conceptually different manner for these two cases respectively. It seems also that the $2n$-dimensional manifolds should be easier to deal with. 

Similar distinction between odd/even dimensions takes place in topological properties of Lie groups i.e.  Betti numbers of Lie group manifolds are given by Betti numbers of certain products of odd spheres \cite{lietop}. Because the structure underlying the regularization of manifolds is expected to be a Lie group it seems plausible that the Lie group cohomology is related to that distinction.

Let us also note that in case of $2n$ dimensions the scaling $N^{2n}$ can be also obtained by considering $N^n\times N^n$ matrices instead of $2n$-arrays. There exist a natural operation,  the Kronecker product of matrices, which gives such matrices from the usual $N \times N$ ones. Therefore there exists a possibility to regularize the $2n$-manifolds by means of usual matrices which are obtained by a Kronecker product of $n$, $N \times N$ ones.   We shall use this observation in Section 4.
 
 \subsection{$d$-brackets}
 For $d$-manifolds the counterpart of membrane APD algebra is the $d$VPD ($d$-volume-preserving diffeomorphism) algebra. The counterpart of the Poisson bracket is the Nambu $d$-bracket \cite{nambubra}
 \begin{equation} \label{nambra}
 \{F_1,\ldots,F_d\}:= \frac{1}{\rho}\epsilon^{i_1 \ldots i_d}\partial_{\varphi_{i_1}} F_1 \ldots \partial_{\varphi_{i_d}}F_d
 \end{equation}
where $\varphi_i$ is the manifold's parameterization, $\rho$ is a certain scalar density to be determined (in this paper we choose $\rho$ as in the case of surfaces i.e. $\rho$ is the square-root of the determinant of the metric of embedded manifold). Technically this is no longer an algebra in a usual sense (as the Nambu $d$-bracket defines a product among $d$ functions and not just $2$ functions) but a different algebraic structure called the $d$-algebra. A natural way to proceed now is to replace the Nambu $d$-bracket by an antisymmetric product of $d$-arrays (modulo some numerical factors e.g. the factor of $-i$ in $\{\cdot,\cdot \} \to -i[\cdot,\cdot]$ for the case of surfaces)
\begin{equation} \label{dcom}
 \{F_1,\ldots,F_d\} \to  c_d[\hat{F}_1,\ldots,\hat{F}_d] : = c_d \epsilon^{i_1 \ldots i_d}\hat{F}_{i_1}\bullet \hat{F}_{i_2} \ldots \bullet \hat{F}_{i_d}
\end{equation}
where $c_d$ is that factor.

Let $Y_A$ be an orthonormal basis of functions ($\int_{\mathcal{M}}Y_AY_B \rho =\delta_{AB}$) on a $d$-manifold $\mathcal{M}$ and consider an expansion of some function $F$ on that manifold
\[
F= \sum_A a_A Y_A.
\]
Although we do not write the range of $A$ explicitly (due to a technical difficulty -  the index $A$ is a multi index and is different for different manifolds), the above sum is over all the modes $Y_A$. Let us assume now that we have a $d$-array like objects $T_A$'s corresponding to $Y_A$'s hence a counterpart of $F$ is
\[
F \ \to \ \ \ \hat{F}:=\sum'_A a_A T_A
\]
where the prime indicates that the sum is over a finite number of modes $\Lambda=\Lambda(N)$ ($\Lambda \propto N^d$ for large $N$).  In order to show that such regularization is indeed correct one has o verify that the structure constants $c_{A_1 \ldots A_{d+1}}$ of the $d$VPD $d$-algebra, given by
\[
\{ Y_{A_1}, \ldots, Y_{A_d} \} = c_{A_1 \ldots A_d A_{d+1}}Y_{A_{d+1}} 
\]
can be approximated by the structure constants $f^{(N)}_{A_1 \ldots A_{d+1}}$ of the regularized $d$-algebra
\[
c_d[ T_{A_1}, \ldots, T_{A_d} ] = f^{(N)}_{A_1 \ldots A_d A_{d+1}}T_{A_{d+1}} 
\]
i.e that $f^{(N)}_{A_1 \ldots A_{d+1}} \to c_{A_1 \ldots A_{d+1}}$. 

In the following section we shall prove that assertion for particular 4-manifolds.

\section{Regularization of $4$-manifolds}

In terms of finding the appropriate regularization for $4$-manifolds, the products of two $2$-manifolds such as $S^2 \times S^2$ or $S^2 \times T^2$ are the simplest ones considering the fact that one knows how to regularize surfaces. From this perspective the manifold $S^1\times S^3$ seems to be much more difficult to deal with. Because the algebra behind surfaces is $SU(N)$ it is reasonable to expect that the algebra underlying regularized 4-manifolds contains two copies of $SU(N)$ as in e.g. the tensor product $SU(N)\otimes SU(N)$. On the other hand, this statement seem highly nontrivial for manifolds such as $S^1 \times S^3$. 

\subsection{The Kronecker product}
Let us start with $S^2 \times S^2$. We shall embed this manifold in $\mathbb{R}^6$ and parametrize its coordinates $(x_1,x_2,x_3,y_1,y_2,y_3)$ as
\[
x_1=\cos \varphi_1, \ \ \ x_2 = \sin \varphi_1 \cos \varphi_2, \ \ \ x_3 = \sin \varphi_1 \sin \varphi_2,
\]
\begin{equation} \label{coord}
y_1=\cos \varphi_3, \ \ \ y_2 = \sin \varphi_3 \cos \varphi_4, \ \ \ y_3 = \sin \varphi_3 \sin \varphi_4.
\end{equation}
Functions on this manifold can be expanded in terms of products of spherical harmonics for each sphere
\[
F(\varphi_1,\varphi_2,\varphi_3,\varphi_4) = \sum_{A_1,A_2}a_{A_1A_2}Y_{A_1}(\varphi_1,\varphi_2)Y_{A_2}(\varphi_3,\varphi_4)
\]
where we again use the double index notation $A_1=(l_1,m_1)$, $A_2=(l_2,m_2)$. If we now replace $Y_A$'s with matrices $T_A$ as in (\ref{tas}) then a question arises what matrix operation is the counterpart of the product $Y_{A_1}(\varphi_1,\varphi_2)Y_{A_2}(\varphi_3,\varphi_4)$. A usual matrix product i.e.
\[
Y_{A_1}(\varphi_1,\varphi_2)Y_{A_2}(\varphi_3,\varphi_4) \ \  \to \ \ T_{A_1}\cdot T_{A_2} 
\] 
is not good. The reason is simple - $T_{A_1}T_{A_2}$ is still $N \times N$ matrix hence the number of independent parameters describing $T_{A_1}T_{A_2}$ scales with $N^2$ while we require the $N^4$ scaling. However the result of the Kronecker product of two matrices $(A\otimes B)_{ijkl}:=A_{ij}B_{kl}$ has the correct scaling, therefore we consider
\begin{equation} \label{map}
Y_{A_1}(\varphi_1,\varphi_2)Y_{A_2}(\varphi_3,\varphi_4) \ \  \to \ \ \gamma_{l_1l_2N} T_{A_1}\otimes T_{A_2}
\end{equation}
where $\gamma_{l_1l_2N}$ is a constant to be determined later on (a counterpart of $\gamma_{lN}$ in (\ref{tas})). That this prescription may work follows form bi-linearity of the Kronecker product and the mixed-product property
\begin{equation} \label{iden}
(A_1 \otimes A_2) \cdot (B_1 \otimes B_2) = (A_1\cdot B_1) \otimes (A_2\cdot B_2)
\end{equation}
which, together with the fact that $T_A$'s are given by linear combination of products of spin matrices $J_i$, implies that we can expand $ T_{A_1}\otimes T_{A_2}$ in terms of (ordinary) products of matrices
\[
J_i^x:=J_i \otimes \mathbf{1}, \ \ \ \ J_i^y:= \mathbf{1} \otimes J_i.
\]
Concretely, we have
\[
 T_{A_1}\otimes T_{A_2}  = \left( \sum_{a_i=1,2,3} c^{m_1}_{a_1 \ldots a_{l_1}}J_{a_1}\ldots J_{a_{l_1}}\right)\otimes   \left( \sum_{b_i=1,2,3} c^{m_2}_{b_1 \ldots b_{l_2}}J_{b_1}\ldots J_{b_{l_2}}\right)  
\]
\begin{equation} \label{iden1}
=\left( \sum_{a_i=1,2,3} c^{m_1}_{a_1 \ldots a_{l_1}}J^x_{a_1}\ldots J^x_{a_{l_1}}\right) \cdot  \left( \sum_{b_i=1,2,3} c^{m_2}_{b_1 \ldots b_{l_2}}J^y_{b_1}\ldots J^y_{b_{l_2}}\right).
\end{equation}

\subsection{The double-Poisson bracket}
Note that the matrices $J_i^x$ and $J_i^y$ form two independent sets of $N^2 \times N^2$ spin matrices i.e.
\[
[J_i^x,J_j^x]=\frac{2i}{\sqrt{N^2-1}}\epsilon_{ijk}J_k^x, \ \ \ \ [J_i^y,J_j^y]=\frac{2i}{\sqrt{N^2-1}}\epsilon_{ijk}J_k^y,
\]
\begin{equation} \label{su2}
[J_i^x,J_j^y]=0
\end{equation}
which  implies that there is no ordering ambiguity in (\ref{iden1}). Moreover let us observe that the commutator $-i[\cdot,\cdot]$ corresponds now to a double-Poisson bracket i.e. defining 
\begin{equation} \label{dpb}
\{f,g\}=\{f,g\}_{\varphi_1\varphi_2} +\{f,g\}_{\varphi_3 \varphi_4}, \ \ \ \ 
\{f,g\}_{\alpha\beta} = \frac{1}{\sin \alpha}\left(\partial_{\alpha} f \partial_{\beta}g- \partial_{\beta}f  \partial_{\alpha}g\right)
\end{equation}
we obtain relations on $S^2 \times S^2$ which are similar to (\ref{su2})
\[
\{x_i,x_j\}=\epsilon_{ijk}x_k, \ \ \ \ \{y_i,y_j\}=\epsilon_{ijk}y_k,
\]
\begin{equation} \label{su2reg}
\{x_i,y_j\}=0.
\end{equation}
Therefore any further construction must take into account the fact that the commutator $-i[\cdot,\cdot]$ is a regularization of the double-Poisson bracket (\ref{dpb}).

\subsection{$4$-arrays}
The advantage of writing everything as in (\ref{iden1}) is that we have got rid of the explicit Kronecker product by "moving" it inside the definition of the building blocks $J_i^x$, $J_i^y$. These matrices can now be manipulated in a usual fashion however since they are $N^2 \times N^2$ the resulting regularization will be much different from that for surfaces. This result is a realization of what was discussed in section 3.2: one can view the Kronecker product $\hat{f} \otimes \hat{g}$ as a 4-array with the indices ranging form $1$ to $N$ 
\[
(\hat{f} \otimes \hat{g})_{i_1i_2j_1j_2}:=\hat{f}_{ i_1i_2} \hat{g}_{j_1j_2}, \ \ \ \ i_1,i_2,j_1,j_2=1,\ldots,N
\]
or as a 2-array (a matrix) with indices from $1$ to $N^2$
\[
(\hat{f} \otimes \hat{g})_{IJ}, \ \ \ \ I,J=1,\ldots,N^2
\]
(we use the notation $\hat{F}$ or $\hat{f}\otimes \hat{g}$ for matrices corresponding to functions $F(\varphi_1,\varphi_2,\varphi_3,\varphi_4)$ or $f (\varphi_1,\varphi_2)g(\varphi_3,\varphi_4)$ on $S^2 \times S^2$).

\subsection{General prescription}
Up to now we have defined a map
\begin{equation} \label{maps2}
x_i \to J_i \otimes \bold{1}, \ \ \ \ y_i \to \bold{1} \otimes J_i
\end{equation}
which gives us the matrix counterparts of the basis $Y_{A_1}Y_{A_2}$ via (\ref{map}) and (\ref{iden1}). However for Cartesian product of arbitrary $2$-manifolds this prescription will not work. On the other hand, we could have obtained the same result by simply defining $Y_{A_1} \to T_{A_1}\otimes \mathbf{1}$, $ Y_{A_2}\to \mathbf{1}\otimes T_{A_2}$
and by requiring that the product of functions $Y_{A_1}Y_{A_2}$ is mapped to a (usual) product of $N^2\times N^2$ matrices  $ (T_{A_1}\otimes \mathbf{1})( \mathbf{1}\otimes T_{A_2})$. This prescription will work since $ (T_{A_1}\otimes \mathbf{1})( \mathbf{1}\otimes T_{A_2})= T_{A_1}\otimes T_{A_2}$ and hence can be used in the general case. 

Let us therefore consider two $2$-manifolds $\mathcal{N}$ and $\mathcal{M}$ and their product $\mathcal{N}\times \mathcal{M}$. Functions on $\mathcal{N}$ and $\mathcal{M}$ can be expanded in terms of modes $Y_{A_1}(\varphi_1,\varphi_2)$ and $\tilde{Y}_{A_2}(\varphi_3,\varphi_4)$ respectively. Suppose their matrix counterparts are $T_{A_1}$ and $\tilde{T}_{A_2}$ and define a map
\begin{equation} \label{genmap}
Y_{A_1}(\varphi_1,\varphi_2) \to T_{A_1}\otimes \mathbf{1}, \ \ \ \  \tilde{Y}_{A_2}(\varphi_3,\varphi_4) \to \mathbf{1}\otimes \tilde{T}_{A_2}.
\end{equation}
Having done that we now define the counterpart of the Nambu $4$-bracket by 
\begin{equation} \label{4nambu}
\{F_1,F_2,F_3,F_4\} \ \ \to \ \ c_4[\hat{F}_1,\hat{F}_2,\hat{F}_3,\hat{F}_4] : =c_4 \epsilon^{ijkl} \hat{F}_i \hat{F}_j \hat{F}_k \hat{F}_l
\end{equation}
where $\hat{F}_i$'s are the matrix counterparts of $F_i$'s and $c_4$ is a number to be determined. The product $\hat{F}_i \hat{F}_j \hat{F}_k \hat{F}_l$ in (\ref{4nambu}) is a usual product of four $N^2 \times N^2$ matrices. In the following we will be interested in calculating the Nambu $4$-bracket for the basis functions $Y_{A_1}\tilde{Y}_{A_2}$ and their matrix counterparts $T_{A_1}\otimes \tilde{T}_{A_2}$ therefore all $F_i$'s considered here will have that product form (and all $\hat{F}_i$'s will have the Kronecker product form).

We now use the observation that the Nambu $4$-bracket can be expanded as \cite{zachos}
\[
\{F_1,F_2,F_3,F_4\} 
\]
\begin{equation} \label{4nambu1}
= \{F_1,F_2\}\{F_3,F_4\} - \{F_1,F_3\}\{F_2,F_4\}+ \{F_1,F_4\}\{F_2,F_3\}
\end{equation}
where $\{\cdot,\cdot\}$ is a double-Poisson bracket, and that the quantum 4-bracket (the 4-commutator) resolves as
\[
[\hat{F}_1,\hat{F}_2,\hat{F}_3,\hat{F}_4]
\]
\begin{equation} \label{4nambu2}
=\{[\hat{F}_1,\hat{F}_2],[\hat{F}_3,\hat{F}_4]\}-\{[\hat{F}_1,\hat{F}_3],[\hat{F}_2,\hat{F}_4]\}+\{[\hat{F}_1,\hat{F}_4],[\hat{F}_2,\hat{F}_3]\}
\end{equation}
i.e. it has twice as many terms due to the noncommutativity. Therefore it is reasonable to expect that $-\frac{1}{2}[\cdot,\cdot,\cdot,\cdot]$ is a correct quantum counterpart of the Nambu $4$-bracket (with the minus factor due to the $i$ factor in $\{\cdot,\cdot \} \to -i [\cdot,\cdot]$ and the fact that the $4$-commutator is a sum of "squares" of commutators, implying that $c_4=-1/2$, c.p. (\ref{4nambu})).
For the case of $S^2\times S^2$ we have shown that the double-Poisson bracket corresponds to the commutator $-i[\cdot,\cdot]$. Let us verify that this rule is true for general product manifolds. Considering the functions in the product form  and their matrix counterparts
\[
F_1:=f_1(\varphi_1,\varphi_2)g_1(\varphi_3,\varphi_4) \to \hat{f}_1\otimes \hat{g}_1 =: \hat{F}_1,
\]
\[
F_2:=f_2(\varphi_1,\varphi_2)g_2(\varphi_3,\varphi_4) \to \hat{f}_2\otimes \hat{g}_2 =: \hat{F}_2
\]
we find that the double-Poisson bracket and the commutator are
\[
\{F_1,F_2\}=\{f_1,f_2\}_{\varphi_1 \varphi_2}g_1g_2 + f_2f_1\{g_1,g_2\}_{\varphi_3\varphi_4},
\]
\[
-i[\hat{F}_1,\hat{F}_2] = (-i[\hat{f}_1,\hat{f}_2])\otimes \hat{g}_1\hat{g}_2 + \hat{f}_2\hat{f}_1 \otimes(-i[\hat{g}_1,\hat{g}_2])
\]
i.e. they resolve in the same way (note the ordering of $f_i$'s). This, together with (\ref{4nambu1}) and (\ref{4nambu2}), verifies that the Nambu $4$-bracket can be replaced by $-\frac{1}{2}[\cdot,\cdot,\cdot,\cdot]$ while keeping in mind that the double-Poisson bracket is replaced by $-i[\cdot,\cdot]$.

\subsection{Leibniz rules}
Let us now note that the Leibniz rule for a double-Poisson bracket i.e.
\[
\{F_1,F_2F_3\} = \{f_1,f_2\}_{\varphi_1 \varphi_2} f_3 g_1g_2g_3 +   f_2\{f_1,f_3\}_{\varphi_1 \varphi_2}g_1g_2g_3   
\]
\begin{equation} \label{leibniz1}
+f_2f_3f_1 \{g_1,g_2\}_{\varphi_3 \varphi_4} g_3+ f_2f_3f_1 g_2\{g_1,g_3\}_{\varphi_3 \varphi_4}.
\end{equation}
resolves in the same way as the commutator
 \[
-i[\hat{F}_1,\hat{F}_2\hat{F}_3] = (-i[\hat{f}_1,\hat{f}_2] \hat{f}_3) \otimes \hat{g}_1\hat{g}_2\hat{g}_3 +   \hat{f}_2(-i[\hat{f}_1,\hat{f}_3])\otimes \hat{g}_1\hat{g}_2\hat{g}_3   
\]
\begin{equation}  \label{leibniz2}
+\hat{f}_2\hat{f}_3\hat{f}_1 \otimes (-i[\hat{g}_1,\hat{g}_2]) \hat{g}_3+ \hat{f}_2\hat{f}_3\hat{f}_1 \otimes \hat{g}_2(-i[\hat{g}_1,\hat{g}_3])
\end{equation}
(where $F_3=f_3(\varphi_1,\varphi_2)g_3(\varphi_3,\varphi_4)  \to \hat{f}_3 \otimes \hat{g}_3:=\hat{F}_3 $). Eqs. (\ref{leibniz1}) and (\ref{leibniz2}) can be used in evaluating $\{F_1,F_2,F_3,F_4F_5\}$ and its quantum counterpart $-\frac{1}{2}[\hat{F}_1,\hat{F}_2,\hat{F}_3,\hat{F}_4\hat{F}_5]$ via identities (\ref{4nambu1}) and (\ref{4nambu2}) respectively. As a result one obtains a Leibniz-like rule, analogous to (\ref{eq2}) for surfaces, which can be used to evaluate the Nambu $4$-bracket and the $4$-commutator for basis functions $Y_{A_1}\tilde{Y}_{A_2}$ and  their matrix counterparts $T_{A_1}\otimes \tilde{T}_{A_2}$ respectively. Just like in the case of $S^2$ we observe that when performing that calculation one makes the same algebraic manipulations modulo factors of $1/N$ (for large $N$) coming from the commutators $[f_i,f_j]$ and $[g_i,g_j]$ (c.p. (\ref{su2}) for the case of $S^2\times S^2$). Therefore in the leading order in $N$ the structure constants $f^{(N)}_{A_1A_2A_3A_4A_5}$ will approximate the $4$VPD structure constants $c_{A_1A_2A_3A_4A_5}$. The large $N$ expression for the factor $\gamma_{l_1l_2N}$ is given by the product $\gamma^{\mathcal{N}}_{l_1N}\gamma^{\mathcal{M}}_{l_2N}$ i.e. of order $N^2$.

Note that identities  (\ref{leibniz1}) and (\ref{leibniz2}) are useful here only because the basis functions on $\mathcal{N}\times \mathcal{M}$ and their matrix counterpart are in the product form and Kronecker product form respectively. However the basis corresponding to e.g. $S^4$ will not have that property and therefore one has to use/argue a different mechanism which contains Leibniz-like rule.  Unfortunately the Leibniz rule for a Nambu $4$-bracket i.e.
\[
 \{F_1,F_2,F_3,F_4F_5\}= \{F_1,F_2,F_3,F_4\}F_5+ F_4\{F_1,F_2,F_3,F_5\}
\]
does not hold for the $4$-commutator. On the other hand, from matrix-regularization point of view one may relax the Leibniz rule to allow $1/N$ corrections -  i.e. the following identity would still work
\begin{equation}  \label{Nleibniz}
 [\hat{F}_1,\hat{F}_2,\hat{F}_3,\hat{F}_4\hat{F}_5]= [\hat{F}_1,\hat{F}_2,\hat{F}_3,\hat{F}_4]\hat{F}_5+ \hat{F}_4[\hat{F}_1,\hat{F}_2,\hat{F}_3,\hat{F}_5] + \hat{O}
\end{equation}
where $\hat{O}$ are some $\hat{F}_i$-dependent terms which vanish in the large $N$ limit. Using (\ref{4nambu2}) and the Leibniz rule for the commutators we find that matrix $\hat{O}$ is
\[
\hat{O} = [F_1,F_4][F_5,[F_2,F_3]]+[F_3,F_4][F_5,[F_1,F_2]]-[F_2,F_4][F_5,[F_1,F_3]]
\] 
\[
+[[F_1,F_2],F_4][F_3,F_5]-[[F_1,F_3],F_4][F_2,F_5]+[[F_2,F_3],F_4][F_1,F_5].
\]
All the terms in $\hat{O}$ contain three commutators hence they are of order $1/N^3$ (assuming that each commutator is of order $1/N$) while the $4$-commutators is of order $1/N^2$ hence we verify that $\hat{O}$ is indeed subleading in $N$.

Therefore in calculating the structure constants one can use the Leibniz rule for $4$-commutator keeping in mind that there will be additional $1/N$ terms. This is another way of proving that the structure constants of $4$VPD are recovered when $N \to \infty$.

\subsection{$S^4$}
To be more specific and show that the finite $N$ Leibniz rule (\ref{Nleibniz}) is required, let us consider a unit $S^4$ embedded in $\mathbb{R}^5$ and given by coordinates
\[
x_1=\cos \varphi_1, \ \ \ x_2 = \sin \varphi_1 \cos \varphi_2, \ \ \ x_3 = \sin \varphi_1 \sin \varphi_2 \cos \varphi_3,
\]
\[
x_4= \sin \varphi_1 \sin \varphi_2 \sin \varphi_3 \cos \varphi_4, \ \ \ x_5 = \sin \varphi_1 \sin \varphi_2 \sin \varphi_3 \sin \varphi_4.
\]
Functions on $S^4$ can be expanded in terms of $S^4$-hyperspherical harmonics $Y^l_{m_1m_2m_3}(\varphi_1,\varphi_2,\varphi_3,\varphi_4)$. These harmonics can be expressed in terms of homogeneous symmetric polynomials in variables  $x_1,x_2,x_3,x_4$ and $x_5$ as
\[
Y^l_{m_1m_2m_3} = \sum_{i_1+i_2+i_3+i_4+i_5=l}a^{m_1m_2m_3}_{i_1i_2i_3i_4i_5}x_1^{i_1}x_2^{i_2}x_3^{i_3}x_4^{i_4}x_5^{i_5} 
\]
\[
=\sum_{a_k=1,2,3,4,5} c^{m_1m_2m_3}_{a_1 \ldots a_l} x_{a_1} \ldots x_{a_l}
\]
where the coefficients $c^{m_1m_2m_3}_{a_1 \ldots a_l} $ are symmetric in indices $a_k$'s. Because the coordinates $x_i$ for $S^4$ do not separate as in (\ref{coord})
a map similar to (\ref{maps2}) will not work. On the other hand, the Nambu $4$-bracket can be evaluated for $S^4$ coordinates, we have
\[
\{x_i,x_j,x_k,x_l\}=\epsilon_{ijklm}x_m, \ \ \ i,j,k,l,m=1,2,3,4,5
\]
where the scalar density used here is $\rho = \sin^3\varphi_1 \sin^2 \varphi_2 \sin \varphi_3$ (c.p. (\ref{nambra})). Therefore a matrix counterpart of $x_i$, call it $\Gamma_i$, should satisfy
\begin{equation} \label{4bras4}
-\frac{1}{2}[\Gamma_i,\Gamma_j,\Gamma_k,\Gamma_l]= d_4\epsilon_{ijklm}\Gamma_m
\end{equation}
where $d_4$ may depend on $N$ ($d_4$ should scale like  $1/N^2$ based on arguments from previous subsection). Here we have assumed that the formula for the quantum Nambu $4$-bracket should be independent of the topology of the manifold (as it is in the case of surfaces) therefore we use (\ref{4nambu}) with $c_4=-1/2$. An obvious choice of matrices satisfying (\ref{4bras4}) is $\Gamma_i \propto \gamma_i$ where $\gamma_i$'s are 5D Dirac matrices corresponding to Euclidean signature $\{\gamma_i,\gamma_j\}=2\delta_{ij}\bold{1}$ - identity (\ref{4bras4}) follows directly from algebraic properties of $\gamma_i$'s. 
However Dirac matrices in 5D are $4\times 4$ and there are no other irreducible representations of a different dimension   while we are looking for $N^2 \times N^2$ matrices $\Gamma_i$ for any $N$. It follows that matrices $\Gamma_i$ cannot satisfy Clifford algebra for $N>2$ while for $N=2$ they are given by 5D gamma matrices.  

The last remark suggests the following construction, define
\begin{equation} \label{gammas4}
\Gamma_i= a\left( \begin{array}{cc}
                      0 & i S_i    \\
                      -iS_i & 0
                              \end{array} \right), \ \ \ 
                              \Gamma_4 = b\left( \begin{array}{cc}
                      0 & \mathbf{1}    \\
                      \mathbf{1} & 0
                              \end{array} \right), \ \ \ 
                              \Gamma_5=b\left( \begin{array}{cc}
                      \mathbf{1} & 0    \\
                      0 & -\mathbf{1}
                              \end{array} \right)
\end{equation}
where $S_i$'s are $n\times n$ spin matrices $[S_i,S_j]=i \epsilon_{ijk}S_k$, $a$ and $b$ are normalization factors so that $\sum \Gamma^2_i=\mathbf{1}$. These matrices are $2n \times 2n$ hence to obtain the correct $N^4$ scaling one should take $2n \propto N^2$ for large $N$. To argue again the importance of the $N^4$ scaling let us also note that since there will be about $2N^4/4!$ matrix counterparts of the harmonics $Y^l_{m_1m_2m_3}$ (c.p. (\ref{scaling}))
\begin{equation} \label{maps4}
Y^l_{m_1m_2m_3} \to T^l_{m_1m_2m_3} := \gamma_{lN}\sum_{a_k=1,2,3,4,5} c^{m_1m_2m_3}_{a_1 \ldots a_l} \Gamma_{a_1} \ldots \Gamma_{a_l}
\end{equation}
the majority of them would be linearly dependent if e.g. $2n \sim N$. The scaling $2n \sim N^2$ guaranties that they are independent. 

For $2n=2$, these matrices are proportional to the 5D  Dirac matrices in  the chiral representation, however for $2n>2$ they no longer satisfy Clifford algebra (e.g. $\{\Gamma_1,\Gamma_2\}\ne 0$ for $2n>2$). On the other hand, as we shall show now, $\Gamma_i$'s satisfy identity (\ref{4bras4}) with 
\begin{equation} \label{abd}
a=\sqrt{\frac{6}{5(n^2-1)}}, \ \ \ b=\frac{1}{\sqrt{5}}, \ \ \ \ d_4=\frac{6}{5}\sqrt{\frac{6}{5(n^2-1)}}.
\end{equation}
For the proof one should a priori calculate twenty four $4$-commutators in (\ref{4bras4}) however due to the complete antisymmetry it is enough to evaluate only 5 of them. Using identity (\ref{4nambu2}) we find that they are
\[
[\Gamma_1,\Gamma_2,\Gamma_3,\Gamma_4]=-4a^3 s(s+1) \Gamma_5, \ \ \ [\Gamma_1,\Gamma_2,\Gamma_3,\Gamma_5]=4a^3 s(s+1) \Gamma_4,
\] 
\[
[\Gamma_1,\Gamma_2,\Gamma_4,\Gamma_5]=-12a b^2 \Gamma_3, \ \ \ [\Gamma_1,\Gamma_3,\Gamma_4,\Gamma_5]=12ab^2  \Gamma_2,
\] 
\[
[\Gamma_2,\Gamma_3,\Gamma_4,\Gamma_5]=-12ab^2  \Gamma_1
\] 
where $s(s+1)$ is an eigenvalue of $n \times n$ matrix representation of the Casimir operator $S^2$ and  $2s=n-1$.
The normalization condition $\sum \Gamma_i^2=\mathbf{1}$ and the fact that $d_4$ in (\ref{4bras4}) should be common to all the $4$-commutators imply equations 
\[
a^2s(s+1)+2b^2=1, \ \ \ -12 a b^2=-4a^3s(s+1)=-2d_4
\]
which are solved by (\ref{abd}) (note that other branches are possible e.g. $a\to -a$, $b\to -b$, $d_4\to -d_4$). Therefore we have found a $2n\times 2n$ irreducible matrix representation of the $4$-algebra (\ref{4bras4}). As anticipated the factor $d_4$ scales like $1/n \sim 1/N^2$ therefore in the large $N$ limit the $4$-commutator is zero which is analogous to the case of $S^2$ (c.p. (\ref{eq1})) i.e. as $N\to \infty$ matrices $\Gamma_i$ $4$-commute. They will not commute however since 
\begin{equation} \label{g45}
[\Gamma_4,\Gamma_5]= \frac{2}{5}\left( \begin{array}{cc}
                      0 & -\mathbf{1}    \\
                      \mathbf{1} & 0
                              \end{array} \right)
\end{equation}
which holds for all $n$.

Having defined matrix counterparts of $Y^l_{m_1m_2m_3}$ by (\ref{maps4}) we can now evaluate corresponding structure constants. Once again the scaling $2n \sim N^2$ turns out to be necessary - otherwise most of the structure constants would be zero (for $2n \sim N$) or there would be too many degrees of freedom (for $2n\sim N^k $, $k>2$). One can now use the Leibniz rule with $1/N$ corrections (\ref{Nleibniz})\footnote{Since the commutator (\ref{g45}) does not scale like $1/N$ one may think that the matrix $\hat{O}$ in (\ref{Nleibniz}) for a generic case may be of order $1/N^2$ (instead of $1/N^3$) and hence not subleading. However $\hat{O}$ will in fact scale like $1/n^2\sim1/N^4$ due to $\Gamma_a$'s, $a=1,2,3$, carrying the factor of $1/n$.} to prove that the structure constants given by $T^l_{m_1m_2m_3}$'s converge to the structure constants of the $4$VPD of $S^4$.
Finally let us observe that form (\ref{dld}) it follows that there are  
\[
\sum_{l=0}^{N-1}D_{l4}=\frac{1}{12}N(N+1)^2 (N+2)
\]
 matrices $T^l_{m_1m_2m_3}$ which for large $N$ is less then $(N^2-1)^2$ hence the group underlying $S^4$ is not $SU(N)\otimes SU(N)$.

\section{$S^3$}
Although in this paper we are mostly interested in $4$-manifolds let us observe that matrices (\ref{gammas4}) can be used to regularize $S^3$. For the coordinates of a unit $S^3$ embedded in $\mathbb{R}^4$  we take
\[
x_1=\cos \varphi_1, \ \ \ x_2 = \sin \varphi_1 \cos \varphi_2, \ \ \ x_3 = \sin \varphi_1 \sin \varphi_2 \cos \varphi_3,
\]
\[
x_4= \sin \varphi_1 \sin \varphi_2 \sin \varphi_3
\]
which satisfy
\begin{equation}  \label{s3bra}
\{x_i,x_j,x_k\}=-\epsilon_{ijkl}x_l, \ \ \ i,j,k,l=1,2,3,4
\end{equation}
with the scalar density $\rho=\sin^2 \varphi_1 \sin \varphi_2$. Just like in the case of $S^4$, in order to find a matrix counterpart of (\ref{s3bra}) it is clear to start with the 4D Dirac matrices $\gamma_i$ owing to the identity $[\gamma_i,\gamma_j,\gamma_k]\propto \epsilon_{ijkl}\gamma_5 \gamma_l$.   However because of the factor of $\gamma_5$ one cannot identify $x_i$'s with $\gamma_i$'s yet. A way to proceed further is to alter the definition of the quantum $3$-bracket and define it using the $4$-commutator as 
\begin{equation} \label{bl}
[\hat{F},\hat{G},\hat{H}]_{\gamma_5}:=[\hat{F},\hat{G},\hat{H},\gamma_5]
\end{equation}
 which therefore gives $[\gamma_i,\gamma_j,\gamma_k]_{\gamma_5}\propto \epsilon_{ijkl} \gamma_l$ and hence $x_i$ can be replaced by $\gamma_i$'s. This approach was used in the context of multiple-membrane theories \cite{bl} where it is also shown that it arises naturally from non-associative algebras. Let us now generalize this construction to $2n \times 2n$ matrices. Using the same $\Gamma_i$'s as for $S^4$ we find that
 \[
[\Gamma_i,\Gamma_j,\Gamma_k]_{\Gamma_5}:=[\Gamma_i,\Gamma_j,\Gamma_k,\Gamma_5] =-2d_4 \epsilon_{ijk5m}\Gamma_m=2d_4\epsilon_{ijkl}\Gamma_l
 \]
therefore one may use these matrices for negative $d_4$ (i.e. the branch  $a\to -a$, $b\to -b$, $d_4\to -d_4$) to represent quantum counterparts of coordinates of $S^3$ as long as we are using $[\cdot,\cdot,\cdot]_{\Gamma_5}$ as a definition of a quantum $3$-bracket. We observe that in the large $n$ limit not only the  $[\Gamma_i,\Gamma_j,\Gamma_k]_{\Gamma_5}$'s  but also the commutators $[\Gamma_i,\Gamma_j]$ are $0$.

The spherical harmonics of $S^3$ can now be mapped to $2n\times 2n $ matrices via. 
\[
Y^l_{m_1m_2}(\varphi_1,\varphi_2,\varphi_3)  = \sum_{a_k=1,2,3,4} c^{m_1m_2}_{a_1 \ldots a_l} x_{a_1} \ldots x_{a_l}
\]
\begin{equation} \label{maps3}
\to T^l_{m_1m_2} := \gamma_{lN}\sum_{a_k=1,2,3,4,} c^{m_1m_2}_{a_1 \ldots a_l} \Gamma_{a_1} \ldots \Gamma_{a_l}
\end{equation}
and therefore to maintain the $N^3$ scaling of the number of modes we will take $2n \sim N^2$. As for $S^4$, matrices $\Gamma_i$ are $N^2\times N^2$ but  the resulting matrix-harmonics (\ref{maps3}) contain only $N^3$ degrees of freedom (more precisely, it follows form (\ref{dld}) that there are $N(N+1)(2N+1)/6$ degrees of freedom corresponding to first $N$ modes). 

That the structure constants coming from $T^l_{m_1m_2}$'s converge to the structure constants of VPD of $S^3$ follows immediately from the fact that the quantum $3$-bracket is in fact given by a $4$-commutator. Therefore the reasoning presented in the case of $S^4$ applies here.

\section{Summary and Outlook}

Matrix regularization of embedded surfaces is a procedure that at first sight seems easily generalizable to $d$-dimensional manifolds. One simply performs Fourier expansion of functions on a manifold and then replaces the modes by a suitably chosen matrices - why should it be more complicated?  The problem however is there - it is much more difficult to find concrete representations of algebraic structures (different from Lie algebras) that should be used to replace the Nambu $d$-bracket. Even more basic problem is related to noncommutative objects that replace functions on the manifold. These objects should depend on roughly $N^d$ real parameters where $N$ is the number of Fourier modes - a natural choice would be to use the $d$-arrays. Although we find this path appealing, in this paper we focused on possible matrix representations due to the ambiguity in defining the product of two $d$-arrays ($d>2$). For $2n$-manifolds we  simply consider $N^n \times N^n$ matrices while for $d=2n-1$ a subset of   $N^n \times N^n$ matrices. 

In this paper we focused on $d=4$ manifolds due to the possible relevance of this case to quantum gravity. By finding a matrix regularization of space-time one may in principle be able to quantize the resulting system of finite degrees of freedom. We firstly considered manifolds that are Cartesian product of two $2$-manifolds and showed that functions on such manifolds can be approximated by $N^2 \times N^2$ matrices from $SU(N) \otimes SU(N)$. This result shows that one does not have to refer to $4$-arrays in order to regularize $4$-manifolds.  We then considered more complicated case: $S^4$, which also can be regularized by $N^2\times N^2$ matrices however this time we were unable to identify the underlying group. Finally we used the construction for $S^4$ to find the regularized description of $S^3$. In particular we used the observation that one can define a quantum $3$-algebra by 
using the quantum $4$-bracket (the $4$-commutator).

Let us now discuss some generalizations. First, as for the product manifolds $\mathcal{N} \times \mathcal{M}$ for arbitrary closed manifolds $\mathcal{N}$, $\mathcal{M}$ of dimension $>1$ we conjecture that the resulting group of matrices underlying the regularization is $G_{\mathcal{N}} \otimes G_{\mathcal{M}}$, where the groups $G_{\mathcal{N}}$, $G_{\mathcal{M}}$ correspond to $\mathcal{N}$ and $\mathcal{M}$ respectively. For the proof, the construction presented in Section 4.4 will work provided the corresponding Leibniz rule (with $1/N$ corrections) is also there. 

Second, the regularization of $S^d$, $d>4$ seems straightforward by appropriate generalization of matrices $\Gamma_i$ (\ref{gammas4}). The strategy is to first regularize $S^{2n}$ with the usual $2n$-commutator as a quantum counterpart of the Nambu $2n$-bracket and then apply this construction for  $S^{2n-1}$. For example $S^6$ embedded in $\mathbb{R}^7$ can be regularized by taking $4n\times 4n$ matrices $\tilde{\Gamma}_i$ 
($i=1,\ldots,7$) with the Nambu $6$-bracket replaced by the $6$-commutator. After this is done the regularization of $S^5$ can be made by using the same matrices (excluding $\tilde{\Gamma}_7$) and defining the quantum $5$-bracket by $[\cdot,\cdot,\cdot,\cdot,\cdot]_{\tilde{\Gamma}_7}=[\cdot,\cdot,\cdot,\cdot,\cdot,\tilde{\Gamma}_7]$.

Third, as observed in \cite{cubic2} there exist $3$-array, $n\times n \times n$ representations of the $3$-bracket for $S^3$. That construction has a virtue of using the usual definition of the quantum $3$-bracket (i.e. the $3$-commutator). Therefore one does not need to refer to the $4$-commutator $[\cdot,\cdot,\cdot,\gamma_5]$. On the other hand, working with $3$-arrays requires introducing a non-standard multiplication rule. It seems  plausible that these two approaches are equivalent.

\section{Acknowledgements} I thank J. Hoppe, W. Li, T. A. McLoughlin and I. V. Melnikov for discussions.
This work was supported by DFG (German Science Foundation) via the SFB grant.

\end{document}